\pdfoutput=1

\documentclass[preprint,journal]{vgtc}       

\ifpdf
  \pdfoutput=1\relax                   
  \usepackage{graphicx}                
  \DeclareGraphicsExtensions{.pdf,.png,.jpg,.jpeg} 
\else
  \usepackage{graphicx}                
  \DeclareGraphicsExtensions{.eps}     
\fi%

\graphicspath{{figures/}{pictures/}{images/}{./}} 

\usepackage{microtype}                 
\PassOptionsToPackage{warn}{textcomp}  
\usepackage{textcomp}                  
\usepackage{mathptmx}                  
\usepackage{times}                     
\usepackage{cite}                      
\usepackage{tabu}                      
\usepackage{booktabs}                  
\usepackage{amsmath}
\usepackage{multirow, makecell}
\usepackage{enumitem}
\usepackage{xspace}
\usepackage[frozencache=true,cachedir=.]{minted}

\usepackage[table,usenames,dvipsnames]{xcolor}
\definecolor{aspHead}{HTML}{2008fe}
\definecolor{aspLiterals}{HTML}{ae3327}
\definecolor{aspParams}{HTML}{1e187c}

\definecolor{aspRed}{HTML}{CD5555}
\definecolor{aspGreen}{HTML}{228B22}
\definecolor{aspBlue}{HTML}{00688B}

\usepackage[ruled,vlined,linesnumbered]{algorithm2e}
\usepackage{amsmath}
\usepackage{MnSymbol}

\SetCommentSty{mycommfont}

\usepackage[T1]{fontenc}

\usepackage{ifthen}
\newboolean{revising}
\setboolean{revising}{true}
\ifthenelse{\boolean{revising}}
{
    \newcommand{\rv}[1]{\textcolor{black}{#1}}
    
} {
    \newcommand{\rv}[1]{#1}
    
}

\usepackage{url} 
\usepackage[hidelinks]{hyperref} 

\newcommand{\etal}{{\textit{et~al.}}\xspace}
\newcommand{\name}{VizLinter\xspace}

\vgtccategory{Research}
\vgtcpapertype{Systems \& Rendering}

\title{VizLinter: A Linter and Fixer Framework for Data Visualization}

\author{Qing Chen, Fuling Sun, Xinyue Xu, Zui Chen, Jiazhe Wang, and Nan Cao}
\authorfooter{
\item
 Qing Chen, Fuling Sun, Xinyue Xu, Zui Chen, and Nan Cao are with Intelligent Big Data Visualization Lab at Tongji University. 
 
 E-mails:
 jane.qing.chen@gmail.com,
 \{fulingsun, xuxinyue\}@tongji.edu.cn,
 \{zuic10a, nan.cao\}@gmail.com.
 Nan Cao is the corresponding author.
\item
 Jiazhe Wang is with Ant Group.
 
 Email: jiazhe.wjz@antgroup.com
}

\abstract{
    Despite the rising popularity of automated visualization tools, existing systems tend to provide direct results which do not always fit the input data or meet visualization requirements. Therefore, additional specification adjustments are still required in real-world use cases. However, manual adjustments are difficult since most users do not necessarily possess adequate skills or visualization knowledge. Even experienced users might create imperfect visualizations that involve chart construction errors. We present a framework, \name, to help users detect flaws and rectify already-built but defective visualizations. The framework consists of two components, (1) a visualization linter, which applies well-recognized principles to inspect the legitimacy of rendered visualizations, and (2) a visualization fixer, which automatically corrects the detected violations according to the linter. We implement the framework into an online editor prototype based on Vega-Lite specifications. To further evaluate the system, we conduct an in-lab user study. The results prove its effectiveness and efficiency in identifying and fixing errors for data visualizations.

} 


\CCScatlist{ 
}

\teaser{
  \centering
  \includegraphics[width=\linewidth]{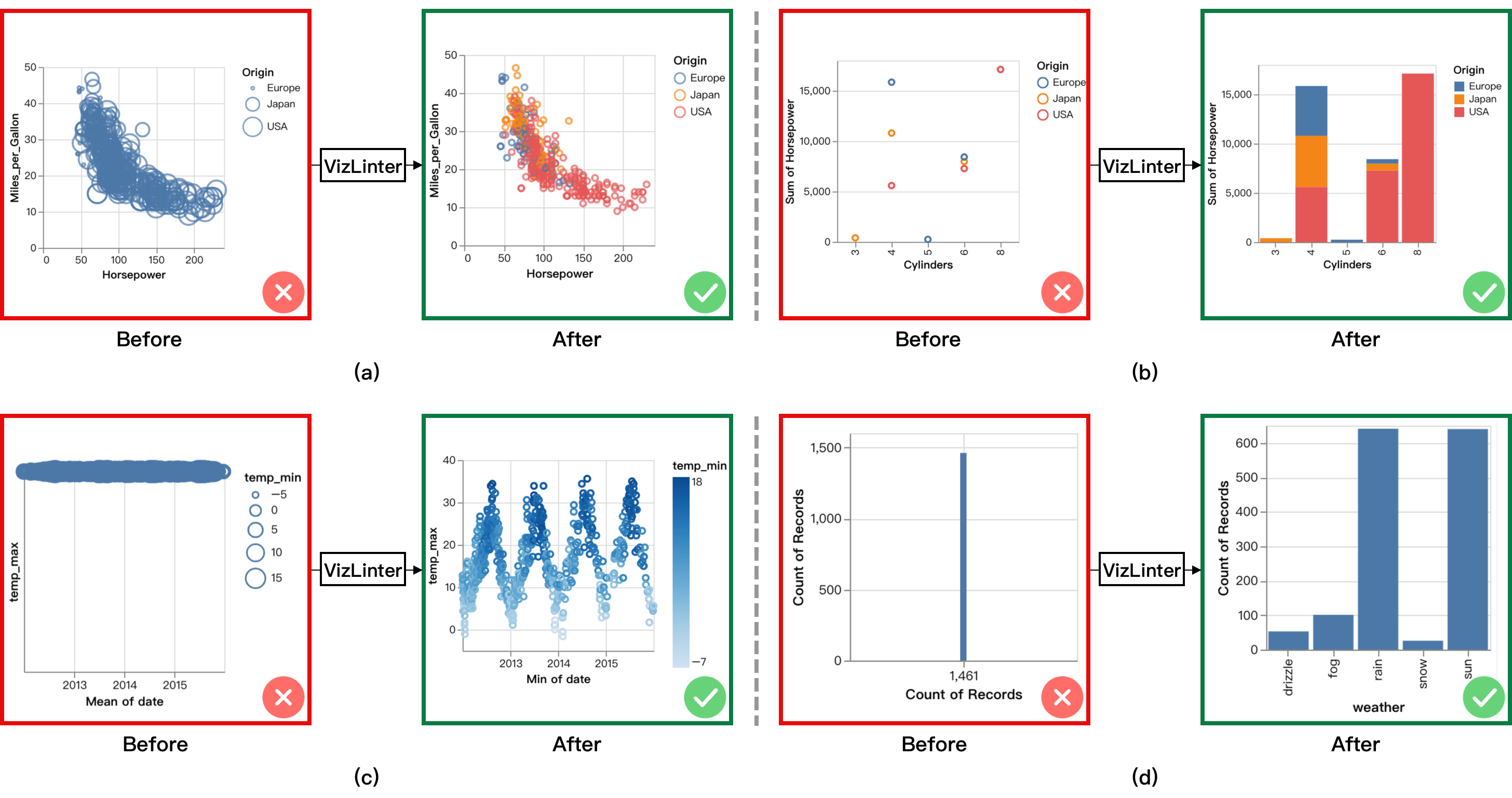}
  \caption{Four example cases of \name. In each case, the chart on the left is the original visualization with errors, and the chart on the right is the one modified by \name.}
  \label{fig:teaser}
}

\begin{document}
%

\firstsection{Introduction}

\maketitle

Visualization tools, such as business intelligence software and programming toolkit, have gained plenty of users as visualization becomes popular across various domains~\cite{kandel_enterprise_2012}. Common visualization tools often require users to set specifications of visualizations either through programming syntax or user interfaces. However, these operations demand users to have solid background knowledge in data analysis and visualization development. The rising demand for creating appropriate visualizations triggers the research interest in developing automated systems to recommend and build expressive and efficient visualizations, especially for users with limited expertise. Most existing automated systems are either rule-based or machine-learning-based. Rule-based systems program visual encoding principles into rule sets to automate the generation of visualizations. They can recommend visualizations based on input data characteristics~\cite{wongsuphasawat_voyager_2016}, intended tasks~\cite{casner_task-analytic_1991}, and user behaviors~\cite{gotz_behavior-driven_2009}. 
Machine-learning-based systems learn relationships between data and the corresponding legitimate visualizations directly~\cite{hu_vizml_2019}. 
Although current recommendation systems are capable of fast-forwarding source data into proper visualizations, they are not always compatible with input data or the visualization requirements due to the limitation of fixed encoding rules and training data. Therefore, manual specifications are inevitable in practice. However, since users of automated systems tend to have inadequate expertise, it is difficult for them to make manual changes while following well-established visual design rationales. 

Although manual adjustments are required when automated results cannot fulfill user needs, few existing works are capable of detecting errors and suggesting corrections for manual inputs. Some prior works are able to highlight problems~\cite{hopkins_visualint_2020,hynes_data_2017} or to lint\rv{, that is, flagging errors in the code} for visualizations~\cite{mcnutt_linting_2018}, but they do not provide any solutions to the detected problems. Other works concentrate on auto-completing user input specifications and recommending optimal alternatives by resolving predefined constraints in an end-to-end manner~\cite{moritz_formalizing_2019,lin_dziban_2020}. However, they do not present any explanations or suggestions on how to fix imperfect visualizations. 

Therefore, programmatic solutions are still absent to detect breached visualization principles and resolve the violations with optimum operations. To remedy this absence, we construct a linter,
\rv{which originally represents static code analysis tools to flag bugs or issues in the programming code,}
to detect violated rules 
\rv{in the visualization}
using Answer Set Programming (ASP)~\cite{lifschitz_what_2008}, a declarative constraint programming language which allows us to model high-level design knowledge into logical facts. 
\rv{An ASP program consists of Prolog-style rules \mintinline[breaklines]{prolog}{a :- L_1, L_2, not L_3, not L_4.}, where
\texttt{\textcolor{aspGreen}{a}}
is the head of the rule and
\texttt{\textcolor{aspBlue}{L\_i}} are literals. The rule states that the head \texttt{\textcolor{aspGreen}{a}} is derived if \texttt{\textcolor{aspBlue}{L\_1}} and \texttt{\textcolor{aspBlue}{L\_2}} are true while \texttt{\textcolor{aspBlue}{L\_3}} and \texttt{\textcolor{aspBlue}{L\_4}} cannot be true. Specifically, the head of each rule in \name is the rule id in the rule base and corresponding parameters, and each literal represents an attribute of the visualization.}
ASP enables easy maintenance and iteration on the rule base~\cite{brewka_answer_2011}, currently refined from the Draco knowledge base~\cite{moritz_formalizing_2019}. 
After violated rules are detected, we formulate a fixer for the broken rules using linear optimization with a designated reward function.

\begin{figure*}
    \includegraphics[width=\textwidth]{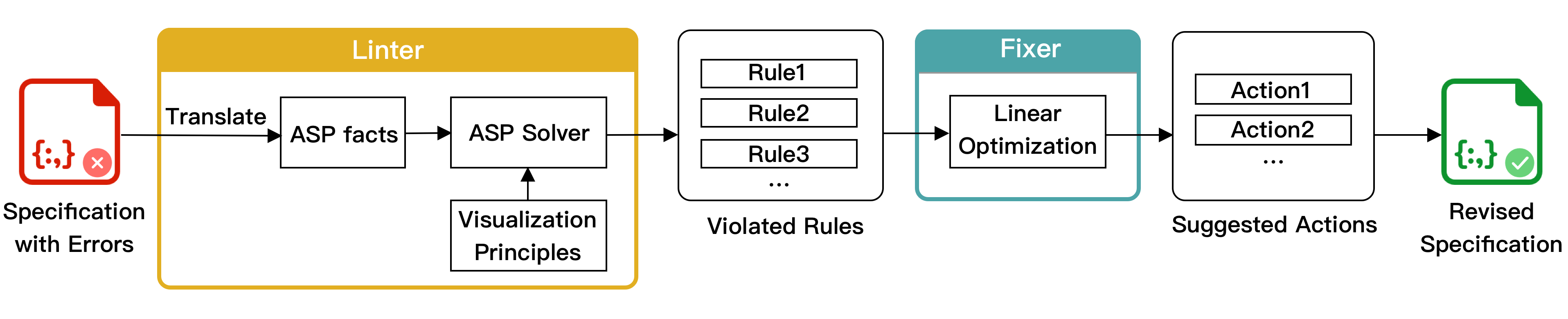}
    \caption{The pipeline of \name includes two key modules: the linter and the fixer. Given the visualization specification with errors, it goes through the linter and the fixer to get the revised specification.}
    \label{fig:pipeline}
    \vspace{-1.5em}
\end{figure*}

In this paper, we propose
\rv{a novel linter and fixer framework for data visualization.}
It detects infringed rules in visualizations and provides solutions to those broken rules by optimizing through linear programming. We implement our framework for Vega-Lite-based visualizations, as Vega-Lite is a widely used language in the visualization community. Our implementation follows the pipeline as shown in Fig.~\ref{fig:pipeline}. It first translates input Vega-Lite specifications into ASP facts. The linter then checks against the facts via an ASP solver and detects the violated principles. Next, the fixer performs the action selected by the optimization algorithm and resolves the broken rules. To demonstrate the usefulness of the proposed framework, we develop a prototype and conduct a user study to assess its efficacy. According to the feedback, our framework is confirmed as helpful by providing ad-hoc prompts and solutions during the creation of visualizations. It can benefit not only business intelligence tool users but also more skilled practitioners and researchers. It is considered as accessible for many existing tools and systems to adapt as an extension and to assist direct development in Vega-Lite coding syntax. 
Besides, the study results indicate the pedagogical value of our framework as users are able to learn visualization specification rules through the interaction.
While the current functionality is already valuable as an automatic error detection and correction tool, it is expected to lint and fix aesthetic or semantic-related issues as well in the next iterations.
To summarize, our contributions are as follows:
\begin{itemize}[noitemsep,topsep=0pt,leftmargin=10pt] 
    \item \textbf{A linter-and-fixer framework.} We present a \rv{novel} linter-and-fixer framework for data visualization, which can examine for erroneous specifications and resolve them automatically.
    \item \textbf{A visualization linter.} We refine the design principle constraints in Draco~\cite{moritz_formalizing_2019} as a concrete rule base and construct a visualization linter to detect violations using ASP. 
    \item \textbf{A visualization fixer.} Using our proposed reward function and the operation transition costs from GraphScape~\cite{kim_graphscape_2017}, we formulate an optimization algorithm using linear programming to fix the violated rules detected by the linter.
    \item \textbf{A user study.} We evaluate the effectiveness of our framework through a user study on the prototype system. Through the study, we validate the effectiveness of our framework in practice and collect instructive reflections for future improvement.
\end{itemize}

\section{Related Work}
In this section, we draw on prior work on recommendation systems, linting, and optimization models for visualization.

\subsection{Recommendation Systems for Visualization}
Visualization recommendation systems accommodate users by performing the heavy-lifting jobs, such as data analysis and insight extraction. They can automatically generate and suggest visualizations for a given dataset. 
Earlier studies tend to recommend visualizations based on predefined rules. They derive rules from statistical features~\cite{vartak_seedb_2015,seo_rank-by-feature_2005,wills_autovis_2010}, visual effectiveness measures~\cite{borkin_what_2013,cleveland_graphical_1984}, and user behaviors~\cite{gotz_behavior-driven_2009} or tasks~\cite{saket_task-based_2018}.
Recent works benefit from machine learning and other algorithms~\cite{wu2021survey}. VizML automatically infers five design choices for input datasets using a machine learning algorithm derived from data characteristics~\cite{hu_vizml_2019}. ClustMe, developed by Abbas~\etal, ranks cluster patterns of monochrome scatterplots using a perceptual visual quality measure to match human judgment~\cite{abbas_clustme_2019}. Shi~\etal present a sequencing model that recommends optimal sequences for different tasks~\cite{shi_task-oriented_2019}. Kim~\etal introduce a directed-graph model for visualization sequence recommendation based on chart similarity~\cite{kim_graphscape_2017}.

Our framework presents a solution to amend visualization with flaws compared to these studies on fast-forwarding visualization recommendation. It allows users to decide if they want to adopt the recommended adjustment and has more pedagogical value. 
Although our framework is also rule-based, we model our rulesets using Answer Set Programming.
Therefore, unlike traditional rule-based recommendation systems whose rule bases are immutable, our framework allows timely updates and modifications on the rules. 

\subsection{Linting for Visualization}
\rv{The idea of linting is borrowed from computer science, which means static code analysis tools to flag bugs or errors in the programming code. Recently, linting has been applied beyond traditional programming languages.}
Before being applied in visualization, the concept of linting has been used in data wrangling as a validation approach to eliminate errors in provided datasets.  CheckCell~\cite{barowy_checkcell_2014}, ExceLint~\cite{barowy_excelint_2018}, and Uni-Detect~\cite{wang_uni-detect_2019} undertake data validation errors in tabular data. Muslu~\etal~\cite{muslu_preventing_2015} and Brachmann~\etal~\cite{brachmann_data_2019} furthermore consider potential problems with data curation and large data sets. Since data wrangling is an important procedure in creating visualizations, works on data linting 
are also valuable for later research on visualization linting.
Meanwhile, in the visualization field, the verification and evaluation of visualizations are also highly relevant to visualization linting. Kirby and Silva emphasize the significance of verifying the accuracy of visualization techniques~\cite{kirby_need_2008}. Prior works undertake the evaluation of visual quality, algorithm performance, and user experience of visualization with different approaches~\cite{isenberg_systematic_2013,gotz_visualization_2019}.
The idea of ``visual linting'' was first introduced by McNutt and Kindlmann to apply linting in visualization~\cite{mcnutt_linting_2018}. They define visualization lint as a framework, that takes in a visualization or its production code, evaluates whether it will pass, and explain why it fails. Their linter targets only readability-related problems in charts, those causing charts difficult for humans to observe~\cite{mcnutt_linting_2018}. Our work, instead, focuses on errors that make charts illegal or prevent them from rendering
\rv{properly, rather than aesthetic problems such as readability issues of a valid visualization.}
After linting being introduced into visualization, some research takes on different perspectives to study the application of linting in visualization. In Metamorphic VegaLite Analyzer, metamorphic testing is applied to validate if a visualization can properly represent the input data~\cite{mcnutt_surfacing_2020}. However, they do not pay special attention to critical errors. Besides, due to reliance on bootstrapping and other statistical techniques, Metamorphic VegaLite Analyzer tends to show limited performance in terms of speed. Since our framework uses Answer Set Programming for rule inspection, which involves less computation, it is able to detect errors without latency.

More recently, Hopkins~\etal design and evaluate VisuaLint, which displays five chart construction errors in visualization using sketchy annotations instead of conventional cues in text messages~\cite{hopkins_visualint_2020}. While the works mentioned above share some similar design considerations with our work, they focus more on the design- and readability-related issues, without any suggestions or functions to help users correct existing errors in the visualization. 

\subsection{Optimization for Visualization}
Visualizations are usually not perfect upon generation due to various factors, such as data volumes or display sizes. Therefore, optimization can help visualizations accommodate different situations and remain good quality. 
Since data quality has a significant impact on the quality of visualization, some research focuses on the data preparation for visualization. For example, Wen and Zhou provide a model to dynamically derive a set of optimal data transformations for the target visualization~\cite{wen_optimization-based_2008}. 
Many other works concentrate on the optimization of charts directly. Some target problems for specific types of charts. Ellis and Dix focus on clutter reduction in parallel coordinates using sampling techniques~\cite{ellis_enabling_2006}.
Trutschl~\etal also tackle point occlusion issues in parallel coordinates but using a self-organized map instead~\cite{trutschl_intelligently_2003}. Wang~\etal improve multi-class scatterplot by optimizing color assignment for class separation~\cite{wang_optimizing_2019}. Heer and Agrawala develop multi-scale banking, enhancing the visual perception of segment orientations in line charts~\cite{heer_multi-scale_2006}. Byron and Wattenberg present a wiggle-minimizing method to optimize the overall shapes of stacked graphs~\cite{byron_stacked_2008}. Tang~\etal develop an authoring tool for storyline visualization, PlotThread, which consists of an AI agent to optimize the layout~\cite{tang_plotthread_2020}. Existing research also covers the optimization for pixel-based charts~\cite{keim_designing_2000}, radial charts~\cite{tatu_combining_2009}, node-link diagrams~\cite{wang_revisiting_2018}, matrices~\cite{behrisch_matrix_2016}, treemaps~\cite{sondag_stable_2018}, geographical maps~\cite{bertini_see_2007}, and text-based charts~\cite{seifert_beauty_2008}. 
Other research resolves general issues regardless of chart types. Draco is a constraint-based system to score visualizations based on the violations against design rules and recommend the top-scored candidate~\cite{moritz_formalizing_2019}.
Lin~\etal contribute Dziban extending on Draco, which additionally optimizes the synthesized charts to accord with the anchor charts~\cite{lin_dziban_2020}.
Wu~\etal address the responsive display issues of SVG-based visualization on mobile screens using reinforcement learning~\cite{wu_mobilevisfixer_2020}. 

However, while previous work could suggest optimal visualizations, they are not able to optimize erroneous visualizations with violated design rules, especially when violations prevent charts from compiling. 
The target of our work is to provide common solutions to various types of visualizations. Specifically, the ideal framework is aimed to improve a given visualization by revising the incorrect specifications. Moreover, to provide the most efficient modification, we adapt the transition cost collected in GraphScape~\cite{kim_graphscape_2017}. A detailed description of the proposed algorithm is presented in Section~\ref{sec:fixer}.

\section{Design Requirements}
In this section, we collect the common issues that practitioners have encountered and summarize the design goals accordingly. 

\subsection{Common Issues in Visualizations}
To better understand the common issues that practitioners encounter and what kind of assistance might be helpful, we conducted semi-structured interviews with two experts. One is an experienced visualization researcher who has \rv{ten-year programming experience} 
and visualization knowledge; the other is the team leader of a data intelligence department in a well-established IT company. Both experts have practical experience in developing data visualization systems \rv{for various domains such as business intelligence and social network analysis.}
In the interviews, the experts noted that creating a visualization requires specifying marks and channels, the two primary components of visual encoding~\cite{bertin_semiology_2010}. However, simply declaring marks and channels will not guarantee a successful encoding. The complex conditions behind it, such as data characteristics, can lead to failed visualizations if not compatible with stated specifications. 
We frame the issues within high-level visualization grammars, such as Tableau/Polaris~\cite{stolte_polaris_2002} and Vega-Lite~\cite{satyanarayan_vega-lite_2017}. These high-level grammars are favored in exploratory visualization due to the preference of conciseness over expressiveness~\cite{heer_interactive_2012}. In high-level visualization grammars, declarations of \textit{marks} and \textit{encoding channels} are required to create visualizations for a given data set. Declaration of \textit{marks} is to specify the mark type (e.g., point or line) for the visualization. Declaration of \textit{encoding channels} includes the specification of the expressed data field (column) and how it is expressed: whether using size or color channel and if any data transformation (e.g., binning, aggregation, logging, scale) or visual transformation (e.g., stacking) are applied. Consequently, four common issues are categorized from the expert interviews and previous literature related to visualization principles~\cite{cleveland_graphical_1984,mackinlay_automating_1986,ware_information_2013,munzner_visualization_2015} and visualization grammar~\cite{wilkinson_grammar_2005}.

    \noindent\textbf{I1.~Incompatibility issues within each encoding channel.} Within-encoding issues illustrate illegal mapping between the selected data field (column) and the encoding channel~\cite{cleveland_graphical_1984,munzner_visualization_2015}. They can lead to failed visualizations when, for instance, a wrong data type is declared, a chosen encoding channel is incompatible with the selected column, or a chosen aggregation is incompatible with the selected column. 

   \noindent\textbf{I2. Incompatibility issues across multiple encoding channels.} Even if each individual encoding is correct, the visualization can still be problematic due to conflicts across encoding channels~\cite{cleveland_graphical_1984,wilkinson_grammar_2005}. One possible cause is that an encoding specification is not compatible with another. It can be attributed to situations such as duplicated usage of the same channel or identical columns used in both axes. 
   
   \noindent\textbf{I3. Incompatibility issues between encoding channels and marks.} Potential failures can occur when the chosen mark is not compatible with the selected encoding channel~\cite{cleveland_graphical_1984,munzner_visualization_2015}, for example, when the size channel is chosen with marks other than points or text. 
    
    \noindent\textbf{I4. Typo issues.} It is natural for humans to make illegal declarations or typo errors during manual input. However, any minor typo mistakes can prevent a visualization from valid rendering. 

According to the above issues, we refine a collection of 41 rules fundamental to construct legitimate charts. \rv{In this paper, we regard a chart as invalid or illegal once it violates one or more rules of the four common issue categories, and possibly fails to render. Illegal chart specifications differentiate from ``readability-related'' problems mentioned by McNutt \etal~\cite{mcnutt_linting_2018} in that they may prevent charts from rendering or presenting correct data. However, ``readability-related'' problems focus on charts rendered successfully with correct data presentation, while other issues, such as a lack of title or annotation or misuse of colors, occur. Such issues are more related to the aesthetics aspect of visualizations, which are excluded in our current work.}
While our current coverage intersects with the one defined by Hopkins~\etal~\cite{hopkins_visualint_2020}, we do not take the issues regarding dual-axis, legends, and ineffective colors into account since the primary focus of this paper is the legibility of visualization. 

\subsection{Design Goals}
\label{sec:design_goals}
Apart from the common issues in visualization, we also interviewed the experts on how they would utilize external support to help find and modify existing issues. Three design goals are thus derived as follows.

\noindent\textbf{G1. Facilitate visualization developers with error indicators.}
Our primary target users are visualization developers with inadequate experience and skilled developers who want to improve their efficiency and accuracy. For those users, concurrent error disclosure could speed up the development process, especially when they are not able to detect problems. Therefore, we need to offer developers the option to see what and where errors are in the specification. In our framework design, this consideration resolves to a linter that inspects errors in visualizations and displays details of the errors.

\noindent\textbf{G2. Automate the fixing operations on detected errors.}
Error indicators alone cannot fully speed up development or ensure correctness because it still takes time and effort to seek solutions for the errors. 
When making manual adjustments, developers may choose operations in a roundabout way. Consequently, we design a fixer to tackle the problem in the most efficient way. We regard the best solution as the one that involves fewer actions while fixing more errors. In Section \ref{sec:fixer}, we present the reward function and the corresponding optimization algorithm based on this logic.

\noindent\textbf{G3. Support easy integration and iteration.}
Many existing business intelligence tools or visualization systems are built on standard visualization grammars. Consequently, we design the liner and the fixer with grammar-based visualization specifications as input. Since Vega-Lite is a standard grammar in visualization research, we use it as the basic data input into our framework. Besides, we choose to form the linter using Answer Set Programming (ASP) so that the ruleset can be easily updated according to different scenarios or domains~\cite{brewka_answer_2011}. Moreover, the fabrication of our linter-and-fixer framework can serve as design guidance for existing tools and systems with similar functionalities.

\section{The \name Framework}

\label{sec:framework}
Guided by the design goals summarized in Section~\ref{sec:design_goals}, we design and illustrate the pipeline of the framework in Fig.~\ref{fig:pipeline}. First, the user inputs a piece of visualization specifications (in our implementation, the inputs are in Vega-Lite JSON format). The framework then transforms the specifications into a list of facts that ASP can handle (\textbf{G3}). Next, the linter built upon ASP examines the facts and detects violations. The corresponding errors are highlighted in the original specifications (\textbf{G1}). For the fixer part, by applying linear optimization programming, it automatically selects an optimal set of operations that resolve the detected violations (\textbf{G2}). Finally, the solutions are translated back to a refined piece of Vega-Lite JSON specifications. 
In this section, we first present the construction of the linter, including the constraints and the answer sets we developed. Then, we describe the algorithm, the reward, and the cost functions used to build the fixer.

\subsection{Visualization Linter}
\label{sec:linter}
The linter is made up of two components, a set of predefined rules visualization should obey and an Answer Set Programming (ASP) solver detecting contravened rules.

\subsubsection{Linting Rules}
A visualization linter, borrowing the concept of ``linting'' from computer science, is a tool built on a collection of rules that should be obeyed. It functions similarly to traditional linting tools in programming, such as ESLint~\cite{eslint}. Like ESLint, \name also examines the violation of rules; however, visualization linting rules include not only syntax errors but also visualization principles. Consequently, the coverage of the ruleset is a determinant of the performance of the linter. The essential principles, which any legal visualization should adhere to, fabricate the basic ruleset for our linter's structure. 
We refine a collection of 41 rules according to the common issues summarized in Section 3.1 and embed them in the linter.

In order to minimize human efforts and potential errors caused by user input, we directly parse data range and cardinality from original data sets. Consequently, some rules related to user input about data properties in Draco are not considered in our collection. Some rules defined in Draco have identical coverage or one's coverage subsets another. In this case, we discard rules with duplicated coverage and keep rules that cover more cases.
Besides, we simplify some complex constraints by breaking them down into more straightforward rules in the fixer. For instance, in Draco, a constraint requiring \textit{stack} to come along with a discrete \textit{color} channel is divided into three distinct rules: one takes care of the absence of color channel when no encoding channel is specified; one deals with the absence of color channel when size is specified as an encoding channel; the third handles when a non-discrete color channel is used. Such decomposition aids in the development of a coherent ruleset and the assignment of resolving operations. Finally, we elaborate a collection of 41 rules as the linter's initial rule base, with a complete list available online\footnote{\url{https://github.com/VizLinter/VizLinter-rules}}.

\subsubsection{ASP for Linting}
We follow the same approach as in Draco to formulate visualization principles using ASP since it enables us to construct a stable model for our rule base. When there is a need to introduce new rules or update the complete ruleset for specific use cases, one can simply add or rewrite the constraints for the intended purposes and not have to re-construct the whole framework again.


\textit{Atoms}, \textit{literals}, and \textit{rules} are the three building blocks for ASP programming~\cite{brewka_answer_2011}. \textit{Atoms} are the elementary propositions that may be true or false; \textit{literals} are atoms A or not A; \textit{rules} are formed by atoms as 
\mintinline[breaklines]{prolog}{a :- L_1, L_2, not L_3, not L_4.}, where
\texttt{\textcolor{aspGreen}{a}}
is the head of these rules and
\texttt{\textcolor{aspBlue}{L\_i}} are literals. 
If all literals are true, then the head \texttt{\textcolor{aspGreen}{a}} is derived; otherwise, \texttt{\textcolor{aspGreen}{a}} is not established. Particularly, if a rule has only the head but no literal, such as \mintinline[breaklines]{prolog}{a :- .}, it is called a fact, indicating that the head is unconditionally true, and such a rule can be abbreviated as \texttt{\textcolor{aspGreen}{a}}. On the contrary, a headless rule, such as \mintinline[breaklines]{prolog}{ :- L_1, L_2.} without head \texttt{\textcolor{aspGreen}{a}} in the rule, is regarded as a constraint, which means \texttt{\textcolor{aspBlue}{L\_1}} and \texttt{\textcolor{aspBlue}{L\_2}} cannot be true at the same time. 

In the linter, we present the refined visualization rules in the \textit{rule} format of ASP. 
For example, the rule \mintinline[breaklines]{prolog}{hard(bin_and_aggregate,C) :- bin(E,_), aggregate(E,_), channel(E,C).} points to encoding \texttt{\textcolor{aspBlue}{E}} with channel \texttt{\textcolor{aspBlue}{C}} using both bin and aggregate. In this rule, a predicate \texttt{\textcolor{aspGreen}{hard}} with two parameters, the rule id \texttt{\textcolor{aspRed}{bin\_and\_aggregate}} and the channel \texttt{\textcolor{aspBlue}{C}}, is used to express the violation. Channel \texttt{\textcolor{aspBlue}{C}} indicates in which encoding channel that the rule is violated. 
Differently, hard constraints in Draco are modeled as a set of headless rules, that is, constraints of ASP, preventing the automatic generation of illegal visualizations that violate these constraints, rather than detecting any violations from the given visualization.


The first step in our linter's process is to translate visualization specifications (Vega-Lite JSON in our case) into ASP programs 
of facts (\textit{rules} without \textit{literals}), characterizing how the given visualization is formed. 
We develop the translator based on Draco to extract the attributes such as mark type, encodings, and their properties. We also incorporate additional functionalities, such as automatically detecting \textit{fieldtype} from data. 
Based on the translated visualization facts and the defined ruleset, we run an ASP solver to find out rules whose literals are all satisfied by the facts, in other words, the errors in the visualization.



\subsection{Visualization Fixer}
\label{sec:fixer}
After acknowledging problems in the visualization, it is still difficult and time-consuming for inexperienced developers to understand the meaning of each linted problem and modify the visualization specification. To better facilitate visualization developers, we propose an optimization algorithm to automatically provide optimal solutions to the detected issues in a given visualization.

\subsubsection{Actions for Rules}
\label{sec:fixer/actions}
\begin{table*}[!t]
\fontsize{8}{8}\selectfont
\caption{Action Space of \name}
\label{tab:actions}

\centering
\renewcommand\arraystretch{1.4}
\linespread{1.5}
\setlength{\tabcolsep}{5mm}{
\begin{tabu}{ p{0.1\textwidth} p{0.2\textwidth} p{0.52\textwidth} } 
\toprule
\textbf{Type}                    & \textbf{Action}    & \textbf{Meaning}                                                                      \\ \hline
\textbf{Mark}                             & CHANGE\_MARK       & Change the mark type for the visualization.                                              \\ \hline
\multirow{7}{*}{\textbf{Encoding}}        & ADD\_CHANNEL       & Add one encoding channel in the visualization.                                        \\
                                 & CHANGE\_CHANNEL    & Change the encoding channel in the visualization.                                     \\
                                 & REMOVE\_CHANNEL    & Remove the encoding channel in the visualization.                                              \\
                                 & ADD\_FIELD         & Add data field in the encoding channel.                                               \\
                                 & CHANGE\_FIELD      & Change the data field used in the encoding channel.                                   \\
                                 & REMOVE\_FIELD      & Remove the data field declared in the encoding channel.                               \\
                                 & CHANGE\_TYPE       & Change the type of the encoding channel.                                              \\ \hline
\multirow{7}{*}{\textbf{Transformation}}       & BIN                & Discretize data values of the encoding channel into a set of bins.                     \\
                                 & REMOVE\_BIN        & Discard data binning in the encoding channel.                                              \\
                                 & AGGREGATE          & Perform aggregation on data values in the encoding channel.                       \\
                                 & CHANGE\_AGGREGATE  & Change aggregation function on the data values in the encoding channel.                        \\
                                 & REMOVE\_AGGREGATE  & Remove aggregation on data values in the encoding channel.                        \\
                                 & STACK              & Stack data values in the encoding channel.                                        \\
                                 & REMOVE\_STACK      & Remove stacking on data values in the encoding channel.                           \\ \hline
\multirow{4}{*}{\textbf{Scale}}           & LOG                & Apply logarithmic transformation on data values in the encoding channel.                 \\
                                 & REMOVE\_LOG        & Remove logarithmic transformation on data values in the encoding channel.              \\
                                 & ZERO               & Require a zero-baseline in the scaled domain of the encoding channel.            \\
                                 & REMOVE\_ZERO       & Waive the zero-baseline requirement in the scaled domain of the encoding channel.              \\ \hline
\multirow{5}{*}{\textbf{Typo Correction}} & CORRECT\_MARK      & Change the illegal mark type to the closest correct one.                 \\
                                 & CORRECT\_CHANNEL   & Change the illegal channel to the closest correct one.           \\
                                 & CORRECT\_TYPE      & Change the illegal data type to the correct one according to data values. \\
                                 & CORRECT\_AGGREGATE & Change the illegal aggregation function to the closest correct one.        \\
                                 & CORRECT\_BIN       & Change the illegal bin number to the correct one.                           \\ \bottomrule
\end{tabu}}
\vspace{-1em}
\end{table*}
To validate the legitimacy, we invited two visualization experts to define feasible actions to solve each rule. In Table~\ref{tab:actions}, all embedded actions are listed with a detailed explanation of what they can do with the visualization.
Because each rule can correspond with several declarations in the specification, breaking the conditions for its establishment is the most effective way to resolve it.
For example, if an encoding misuses the log scale, one can remove the log scale in the encoding or even change the encoded data field to another quantitative data field. Following such methodology, each rule has at least one feasible action and a maximum of five actions, where a rule is established corresponding to several different situations. 

\subsubsection{Score of Action}
In single-problem cases, it is straightforward to find a solution. However, in situations where there are multiple problems, potential solutions are rarely unique. To find the optimal action sets to resolve the violated rules, we construct an optimization algorithm that considers the reward and the cost of actions. 

We define each action's reward as its contribution to the visualization, that is, how many problems in the visualization can be solved after completing it. The reward is composed of two parts. First, we evaluate the proportion of problems solved in the visualization:
\begin{equation}
    reward^+(a) = \frac{|R_i - R_{i+1}|}{|R_i|}
\end{equation}
where $R_i$ and $R_{i+1}$ represent the violated rulesets of visualization before and after the specific action $a$. Here we use the relative complement of $R_i$ with respect to $R_{i+1}$ to represent the solved problems.
However, there are occasions where such actions can bring unintended consequences, one of which is violating new rules.  
For example, a visualization violating the rule that x-axis and y-axis cannot perform count aggregation simultaneously has two possible actions, removing count aggregation from the x-axis or y-axis. However, suppose the channel where the aggregation is removed happens to have no declared data field. A newly violated rule will then derive that an encoding channel should declare the data field or use count aggregation.
\begin{equation}
    reward^-(a) = \frac{|R_{i+1} - R_i|}{|R_{i+1}|}
\end{equation}
In this case, we punish the action $a$ by the proportion discrepancy between $R_{i+1}$ and $R_i$ to show the new violated rules caused by it. Hence, the overall reward of an action $a$ is computed by subtracting its input from its side effect. To concentrate on the positive effect of an action, we set the default weight $w = 0.05$ to tolerate the side effects brought by an action. 
\begin{equation}
    reward(a) = reward^+(a) - w \times reward^-(a)
\end{equation}
When making corrections, actions modify visualizations in different ways.
The transition cost for the action, adapted from GraphScape~\cite{kim_graphscape_2017}, is defined to model the changes between visualization $V_i$ and $V_{i+1}$ resulting from the action. If multiple actions can solve an error in the visualization, then the action with the lowest transition cost is the best solution for the problem. 
\begin{equation}
    cost(a) = transition\_cost(V_i, V_{i+1})
\end{equation}
The overall score of an action is measured by its reward and cost. After trials during experiments, we set the default weight $\alpha = 0.8, \beta = 0.2$ to encourage the following algorithm to pay more attention to the benefits of actions: 
\begin{equation}
    score(a) = \alpha \times reward(a) - \beta \times cost(a)
\label{eq:score}
\end{equation}
\subsubsection{Optimization Algorithm}
After calculating the scores of all candidate actions for rectifying a given visualization, the next problem to consider is which sequence and combination of actions can fix the errors with optimal scores. 

Assume that a visualization has $n$ broken rules $r_1, r_2, \dots, r_n$. Each rule $r_i$ has $m_i$ candidate actions $a_{i,1}, a_{i,2}, \dots, a_{i,m_i}$, implying that this rule has $m_i$ possible solutions. 
Meanwhile, each action $a_{i,j}$ has its corresponding score calculated by Formula~\ref{eq:score}. 
In the algorithm, we set the candidate action $a_{i,j}$ as a binary variable to indicate whether to select it eventually. 


Therefore, our problem can be converted into a Binary Integer Programming(BIP) problem as:
\begin{subequations}
\begin{alignat}{3}
  & \text{maximize}   & \quad & \displaystyle\sum_i^n \sum_j^{m_i} score(a_{i,j}) \times a_{i,j}     \label{eq:obj1}  \\
  & \text{subject to} &       & \sum_{j=1}^{m_i} a_{i, j} = 1,\quad i \in \{1, \dots, n\} & \label{eq:cons1}\\
  &                   &       & a_{i,j} = a_{x,y}, \quad \text{if $a_{i,j}$ and $a_{x,y}$ are equivalent actions} \label{eq:cons2}\\
  &                   &       & a_{i,j} \in \{0, 1\},  \quad
                                                 \begin{tabular}[c]{@{}l@{}l}
                                                 $i$ &$\in$ $\{1, \dots, n\} $ \\
                                                 $j$ &$\in$ $\{1, \dots, m_i\} $
                                                 \end{tabular}  & \label{eq:cons3}
\end{alignat}
\end{subequations}
The above programming objective is to maximize the accumulative score of all candidate actions taken to fix the flawed visualization. 
To prevent redundant operations, Constraints~\ref{eq:cons1} guarantee that one rule is fixed by only one of its candidate actions; therefore, the sum of the assigned value of actions from one rule can only be one. 
Since each rule's candidate actions are defined respectively as mentioned in Sec.~\ref{sec:fixer/actions}, there may be equivalent actions in different rules, that is, actions with identical names (e.g., REMOVE\_AGGREGATE) performing in the same encoding (e.g., x). 
Constraints~\ref{eq:cons2} ensure that any two equivalent candidate actions selected by different rules should be assigned with the same value, i.e., picking up identical actions simultaneously or discarding both. It is used to eliminate situations where a rule $r_2$ uses a different solution when the problem could have been solved by using the adopted action of another rule $r_1$.
Constraints~\ref{eq:cons3} restrict values of all actions $a_{i,j}$ to be either 0 or 1, indicating the status of not being selected or being selected respectively. Hence, the objective function~\ref{eq:obj1} is determined solely by the scores of the adopted actions.

\setlength{\textfloatsep}{5pt}
\begin{algorithm}[!tb]
\label{alg:update}
\SetAlgoLined
\SetKwInOut{Input}{Input}
\SetKwInOut{Output}{Output}
\Input{action $\mathcal{\alpha}$, original violated ruleset $\mathcal{R}_i$, violated ruleset $\mathcal{R}_{i+1}$ after performing action $\mathcal{\alpha}$, candidate actions $\mathcal{A}$ of each rule}
\Output{updated $\mathcal{A}$}
\tcc{Traverse each rule $r$ that action $\mathcal{\alpha}$ has solved}
\For{$r$ {\upshape \textbf{ in }} $\mathcal{R}_i - \mathcal{R}_{i+1}$}{
    \tcc{Retrieve candidate action list of rule $r$}
    A $\leftarrow$ $\mathcal{A}$($r$)\;
    \If{$\mathcal{\alpha}$ $\not\in$ A}{
        update $\mathcal{A}$ by adding $\mathcal{\alpha}$ in A\;
    }
}
\Return $\mathcal{A}$\;
\caption{Update Candidate Actions of Each Rule}
\end{algorithm}

By providing a list of candidate actions, we address the above BIP problem by finding the optimal action set with the highest accumulative scores. BIP is an NP-Complete problem, whose solution can be found using the Branch-and-Cut algorithm~\cite{padberg1991branch} or various well-established solvers such as SCIP~\cite{achterberg2009scip} and CBC~\cite{forrest2005cbc}.
We utilize the Python package PuLP~\cite{mitchell2011pulp} to solve the BIP problem in our implementation. PuLP is a widely-used linear programming API for defining problems and invoking multiple external solvers. We conduct all optimization using Python 3.7 with PuLP 2.4 with the default CBC solver on a computer equipped with 2.4 GHz Intel Core i5 and 16GB RAM. 
All the testing cases for our problem can be solved within seconds.


The above optimization has a relatively good performance, especially when all the violated rules in the visualization and their corresponding candidate actions are independent. 
However, in some testing cases, candidate action $a_\alpha$ of rule $r_\alpha$ helped to fix the issue brought by rule $r_\beta$, which was also solved by action $a_\beta$ adopted by the optimization algorithm.
When defining each rule's feasible actions as its possible solutions, the experts only consider the rule itself, regardless of the co-occurrence of other rules. Therefore, cases occur when a rule is solved by a non-predetermined action. To address such an issue, we update constraints~\ref{eq:cons1} by expanding candidate actions of Rule $r_i$ with the actions of other rules that also contribute to solving $r_i$ as described in Algorithm~\ref{alg:update}. We then solve the BIP model with the updated constraints. In the optimization result, no more redundant actions are recommended, providing users with simple and straightforward fixing suggestions.

Fig.~\ref{fig:teaser} shows four examples of how \name detects issues in the visualization (shown as Before) and then corrects them (shown as After).
The data used in the examples are from the sample dataset of Vega-Lite\footnote{\url{https://vega.github.io/vega-datasets/}}.
Fig.~\ref{fig:teaser}(a) and Fig.~\ref{fig:teaser}(b) are built based on the \textit{car} dataset, including records of cars and their basic properties.   
The original visualization of Fig.~\ref{fig:teaser}(a-Before) depicts the relationship of \texttt{Horsepower}, \texttt{Miles\_per\_Gallon} of cars and their \texttt{Origin}, where the size channel is not compatible with the nominal data field \texttt{Origin} (\textbf{I1}). \name rectifies it by changing the size channel to the color channel as shown in Fig.~\ref{fig:teaser}(a-After). 
One rule of \textbf{I3} type is violated in Fig.~\ref{fig:teaser}(b-Before), where mark type point is not suitable to depict data with stacking. Changing the chart type by modifying mark type to bar corrects the visualization, recommended by \name.
Fig.~\ref{fig:teaser}(c) and Fig.~\ref{fig:teaser}(d) visualize data about daily weather reports from the \textit{seattle-weather} dataset.
Fig.~\ref{fig:teaser}(c) before modification misuses the size channel with the data field \texttt{temp\_min} containing negative values (\textbf{I1}) and performs log transformation incorrectly in the y-axis encoded the data field \texttt{temp\_max} holding negative values (\textbf{I1}). \name removes the log transformation in the y-axis and substitutes the size channel with the color channel. 
For the visualization of Fig.~\ref{fig:teaser}(d), both x-axis and y-axis execute count aggregation. As a result, only a single bar is visualized (\textbf{I2}). \name discards the aggregation on the x-axis to represent the number of appearances of different weather types.

\section{Implementation}

Following the framework described above, we develop a Python package, \texttt{vega-lite-linter}\footnote{\url{http://vegalite-linter.idvxlab.com/}}, which embeds a function \texttt{lint()} to detect errors in Vega-Lite syntax and a function \texttt{fix()} to provide the optimal solutions with the fewest steps. By automatically detecting and resolving errors, \texttt{vega-lite-linter} enables visualization developers to build accurate charts quickly in Python. Meanwhile, \texttt{fix()} provides alternative actions that can resolve each rule and ranks them in the order of scores calculated by Formula~\ref{eq:score}. Users can refer to this action list and choose the solutions to fulfill their own needs.



\label{sec:prototype}

\begin{figure*}[!t]
    \includegraphics[width=0.88\textwidth]{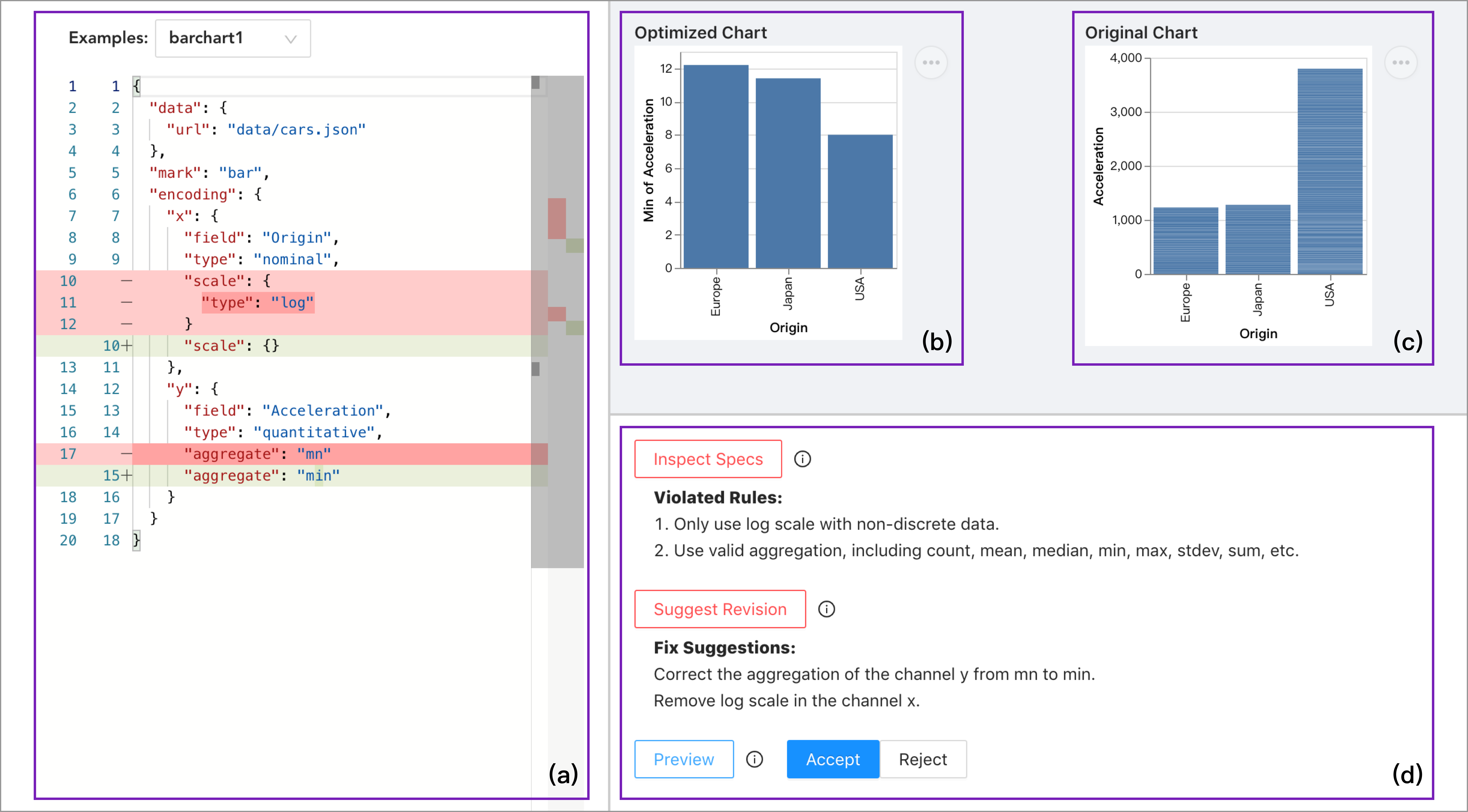}
    \centering
    \vspace*{-2mm}
    \caption{The interface of \name's web prototype consists of four views: (a) the code editor with highlighted differences from the original Vega-Lite JSON, (b) the optimized chart after revision, (c) the original chart before revision, and (d) the toggles (and detailed messages) of the linter, the fixer and its modification preview.}
    \label{fig:prototype}
    \vspace{-1.5em}
\end{figure*}

Based on the above package, we then create a web prototype\footnote{\url{http://vizlinter.idvxlab.com/}} with the Python backend, as shown in Fig.~\ref{fig:prototype}, so that it is easily accessible to any Vega-Lite developer to build and validate visualizations. 
We customize the interface to better serve the framework, referring to the original Vega-Lite Online Editor~\cite{satyanarayan_vega-lite_2017}. 
In addition to the basic code editing panel and chart render panel, we integrate a prompt panel for the linting and fixing functions of \name as shown in Fig.~\ref{fig:prototype}(d). 
When clicking the ``Inspect Specs'' button, any violated rules of the current visualization specification will be shown. Users can correct the specification on their own according to the description of violated rules or check our suggested actions by clicking the ``Suggest Revision'' button. 
Users can click the ``Preview'' button to preview the automatic modification result. The corresponding revised specification will show up in the code panel. ``Accept'' and ``Reject'' buttons are provided for users to approve then apply or reject the revisions. Meanwhile, the chart panel will render the visualization with the revised specification. 

\section{Evaluation}

        
\subsection{User Study}
The user study examined the effectiveness of the proposed framework. \rv{Specifically, we invited the participants to use \name and asked whether they agreed with the issues identified by \name and the corresponding solutions. After the study, we analyzed and discussed both the quantitative and qualitative results.}

\subsubsection{Participants}
We invited 20 participants with visualization experience to take part in our study. Twelve of them are from a technology and financial services company (four females and eight males), whose daily work involves developing visualizations with business intelligence tools. The other 8 participants are visualization researchers (six females and two males). The participants' ages range from 23 to 30 ($M=25.6, SD=2.0$).

\subsubsection{Preparation}

We prepared 15 flawed visualizations in Vega-Lite JSON format using the three sample datasets from Vega-lite, including \textit{cars}, \textit{airports}, and \textit{seattle-weather}. Among these visualizations, five included one error, five two errors, and the rest contained three errors.
There were 25 distinct rules involved in these cases, covering 61\% of those in our framework's rule base. 



We developed a user study system adapted from the \name prototype, whose interface was split into two panels to enable direct comparisons. The left panel presented the original Vega-Lite specification with flawed visualization in a read-only mode. The editable code panel on the right was provided for participants to revise and edit on their own, with the corresponding visualization rendered simultaneously. After participants finished revising each question, they were encouraged to click ``Inspect Specs'' and ``Suggest Revision'' to view the linter detected violations and the fixer's suggestion by \name. If participants agreed on the suggestions, they could click ``Accept Suggestion'' to apply the advised actions on the original visualization automatically. After accepting the fixer's suggestion, participants were still allowed to edit the revised visualization further.

\subsubsection{Procedure}

The study started with a \rv{twenty-minute} training session about Vega-Lite grammars. We also introduced the four categories of common issues (\textbf{I1} $\sim$ \textbf{I4}) within visualizations and a demo case of the upcoming tasks. To verify they understood the instructions, we asked them to try the test questions and answered their questions regarding the user study. Each participant was then given 15 flawed visualizations with the original Vega-Lite specifications. The participants were asked to find out the errors and fix them according to their visualization knowledge and experience. Their revisions in the Vega-Lite specifications were highlighted in green on the editable code panel, and the mutated lines in the original specification were highlighted in red on the original code panel. When the participants completed the task or felt stuck, they were asked to use the linter and the fixer functions of our system. Two questions were then asked: (1) whether they acknowledged the detected errors by the linter and the proposed solutions by the fixer, and (2) whether they would like to make further edits on the improved chart.
After each participant finished all the tasks, we conducted a brief interview to collect feedback on general user experience, effectiveness, and suggestions for our \name framework.

\subsection{Analysis \& Results}

We obtained participants' revisions and their acceptance decisions for each suggestion provided by \name, as well as their completion time of each question and the correction rate. We first analyzed the effectiveness and efficiency of our framework by computing the completion time, the correction rate, and the acceptance rate. Then, we summarized their feedback from interviews to further evaluate the prototype system.


\label{sec:analysis}
\subsubsection{Quantitative Analysis}
\begin{figure*}[!t]
    \includegraphics[width=\textwidth]{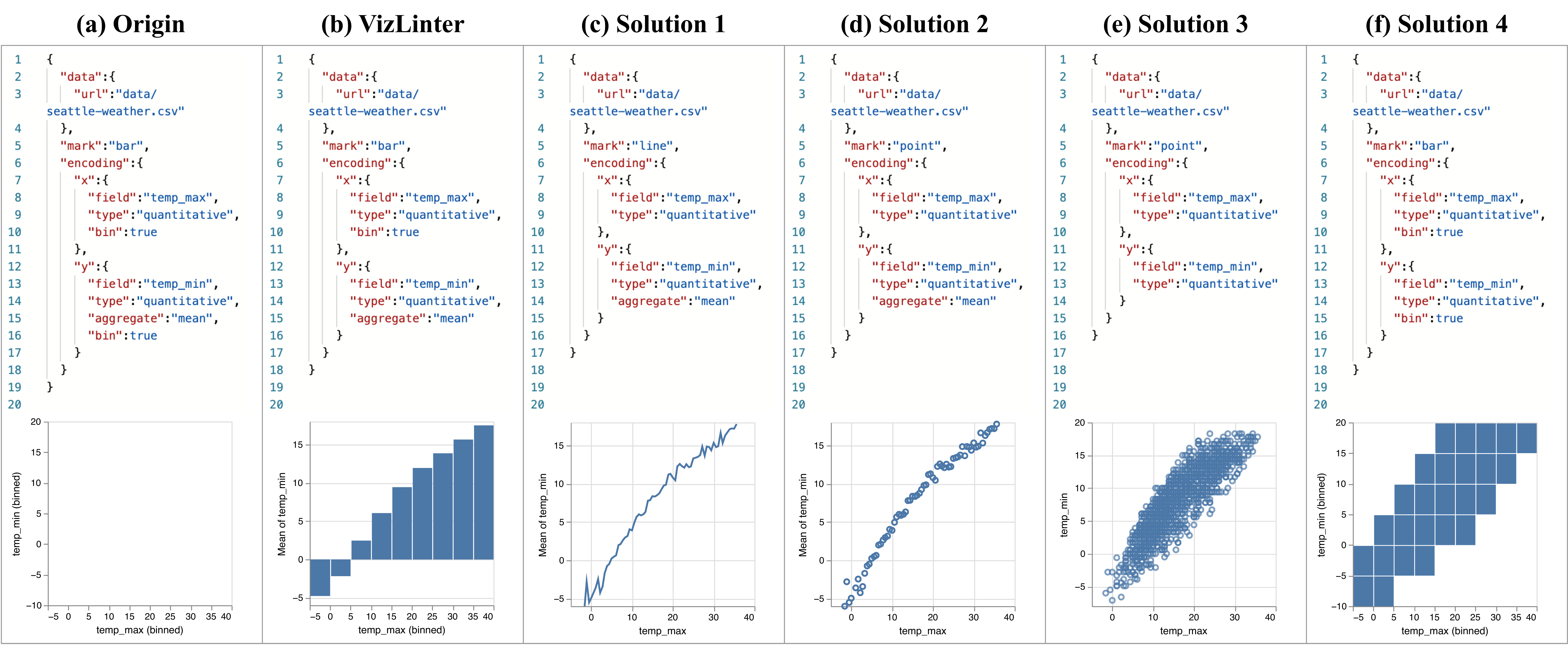}
    \vspace*{-7mm}
    \caption{One case from the user study: (a) the original visualization with errors; (b) the suggested modification result made by \name; (c),(d),(e),(f) four modificated results by different participants.}
    \label{fig:case}
    \vspace{-1.3em}
\end{figure*}
\paragraph{\textit{Completion Time.}} 
It took the participants 97.7 seconds on average to find and fix errors in each visualization ($SD=28.1$). 
The corresponding mean completion time for correction of the different number of errors (one, two, and three) in the visualization was 101.3 seconds ($SD=23.0$), 93.8 seconds ($SD=27.9$), and 98.1 seconds ($SD=32.1$), respectively. 
Contrary to our expectations, the error fixing time did not increase with the number of errors alone. Instead, it was related to the difficulty of figuring out solutions and the complexity of manipulating the specification as well. For example, adding a new encoding channel resulted in more steps than simply modifying the aggregation function within an encoding. 


\paragraph{\textit{Correction Rate.}} We inspected the edited specifications and recorded the number of remaining issues after user revision to calculate the correction rate. The average correction for all the questions was 77\% ($SD=19.7\%$).
We also calculated the average correction rate of four types of common issues. Rules related to issues within each encoding channel (\textbf{I1}) had a 75.2\%  ($SD=22.7\%$) correction rate, while the average correction rate of those related to issues across multiple encoding channels (\textbf{I2}) was 70.8\%  ($SD=20.8\%$). 
Rules related to issues between encoding channels and marks (\textbf{I3}) gained an average correction rate of 82.5\% ($SD=14.4\%$), and typo issues  (\textbf{I4}) reached 90\% ($SD=5\%$) in average correction rate. The latter two types of errors in \name seemed easier to resolve than the first two, potentially because some details inside encoding channels were more difficult to be noticed or fixed.

We summarized some cases that the participants failed to correct and analyzed their causes. A typical case is when participants did not notice the error and could not figure out feasible solutions based on their knowledge. For example, Question 2 was corrected by only four participants, where a rule \textit{``Use at most 20 categorical colors in the visualization''} was violated. Some participants did not recognize this problem due to unawareness of the corresponding design principle or habitual neglect of color overuse in their daily practice. While some participants noticed this error, they still failed to resolve it. This is due to the difficulty of manipulating encoding channels, as optimizing color encoding is also an important research topic for effective visualization design~\cite{wang_optimizing_2019}. Another rule, \textit{``Channel size is not suitable for data with negative values,''} involved in Question 8 and Question 14, also led to low correction rates because only some participants realized that it was illegal to encode size channel for negative data.

Among the rules embedded in \name, some violations do not intervene with visualization rendering due to built-in exception handling of Vega and Vega-Lite.
One example is shown in Fig.~\ref{fig:prototype}, where two rules \textit{``use log scale with non-discrete data''} and \textit{``use valid aggregation, including count, mean, min, etc.''}
were violated, but the visualization rendering was not affected by the invalid log scale. Such errors could only be found in the codes and sometimes were ignored by the participants. We believe that specifications with unnecessary declarations would be error-prone in creating visualizations. Hence, \name can facilitate users to identify errors that are not obviously reflected in the rendered visualizations. 


\paragraph{\textit{Acceptance Rate.}} 
The average acceptance rate was 90\% ($SD=14.2\%$), indicating that most of the suggestions were adopted.
For those recommended solutions with a low acceptance rate, there were four types of scenarios.
First, some violated rules were questioned by a few participants. For example, some participants disagreed with the rule requiring the y-axis to start from zero in bar charts. Consequently, no action was taken for this violation.
Second, after comparing the revision by \name and by themselves, some participants favored their own revisions. For example, one case shown in Fig.~\ref{fig:case}(a) involved two violated rules: \textit{``Mark bar, tick, line, and area require some continuous variables on x- or y-axis''} and \textit{``Use both binning and aggregation on the data at the same time is illegal.''} \name's one-step fixer suggestion was to remove the bin operation on the y-axis shown in Fig.~\ref{fig:case}(b). While some participants made the same choice as \name, others attempted different solutions and rejected the suggestion provided by \name. A common alternative was to remove the bin operation of both x- and y-axis and to change the mark type from bar to line (Fig.~\ref{fig:case}(c)) or point (Fig.~\ref{fig:case}(d)), then to discard the aggregation on the y-axis to depict the distribution (Fig.~\ref{fig:case}(e)). Another interesting result from one developer was to ignore the former rule and only remove the aggregation on the x-axis, as shown in Fig.~\ref{fig:case}(f). \textit{``I did see such kind of chart at work,''} explained the participant. 
\rv{For the same visualization with flaws, \name and the participants sometimes gave different solutions for revising the specification to correct the chart. Since the participants' revisions are influenced by their preferences in visualization, it is difficult to evaluate these subjective choices. However, in future studies, it might be possible to assess the revisions based on design considerations such as aesthetics.}
Third, some recommended actions were regarded as neglecting user intents, therefore rejected by some participants. For example, in Question 12, the encoding specification broke three rules by binning and aggregating simultaneously on nominal data: \textit{``Only use binning on quantitative or ordinal data,''} \textit{``Nominal data cannot be aggregated,''} and \textit{``Use both binning and aggregation on the data at the same time is illegal.''} The fixer recommended REMOVE\_AGGREGATE and CHANGE\_FIELD to correct the violations. On the contrary, the participants took the REMOVE\_AGGREGATE and REMOVE\_BIN actions to fix the errors to avoid changing the original data mapping. However, this combination broke the framework's constraint that only one candidate action for each rule could be selected since REMOVE\_AGGREGATE and REMOVE\_BIN were both candidates for the rule \textit{``Use both binning and aggregation on the data at the same time is illegal.''}
This case indicates one limitation of \name as it takes only one candidate action for a single rule other than combinational actions.
The last cause for rejection was that some participants believed there were other solutions outside of the action scope to fix the problem. For example, the suggested revision for Question 2 mentioned above was to remove the whole color encoding since it was not suitable to describe the nominal data over 20 categories. Though it was the best option within the current scope, some participants argued the chart after applying this action was still not good enough, even if it did meet the basic design principles, \textit{``I expect another chart type or more aggregation on the color channel instead of deleting it.''}
After reviewing the edited specifications, we discovered new errors after human revisions. While some revisions partially or fully correct the original problems, new errors were introduced into the visualizations. 
Twenty-three of all the 300 chart modifications ($15\text{ questions}\times20\text{ participants}$) violated rules that did not appear in the original problems. This result illustrates our \name framework's advantage over manual adjustments in that it prevents new problems after modifications to the original specifications.

\subsubsection{Qualitative Analysis}

To further evaluate the usability of \name, we interviewed the participants after the completion of all the tasks. 


{\large \textit{Overall Performance.}}
All the participants agreed that \name was helpful as it could automatically find errors in visualizations and provide suggestions to fix them. One participant commented, \textit{``for those who are not familiar with the visualization guidelines, it is challenging to make a legitimate visualization on their own. Tools as \name would really help a lot.''}
In terms of efficiency, all the participants responded that \name saved a lot of time and made it possible to revise a chart with a single click. They appreciated its efficiency and commented, \textit{``this really helps me save time, especially when I attempt to create charts with unfamiliar datasets.''}

{\large \textit{Instructive Value.}}
Some participants mentioned that \name could help quickly learn visual design guidelines during the linting and fixing process. One participant said, \textit{``when I employ this tool, I can pick up a lot of visualization principles which I used to neglect.''}
Another participant noted, \textit{``this tool expands my ideas, helps me think out of the box and offers some options for modification.''}

{\large \textit{Suggestions.}}
In spite of the above positive feedback, we also summarize and discuss the suggestions from the participants. 

\begin{itemize}[noitemsep,topsep=0pt,leftmargin=10pt]
    \item {Consider user intent.} 
    Recommended solutions for visualizations by \name usually result in minimal changes, but they do not always match users' expectations. When the number of revision actions is not a concern, users often have different ideas for making corrections.
    More than half of the participants pointed out that user intent should be considered.
    One participant said, \textit{``I would like to see the tool taking more user intentions into account and suggest more revision options to choose from.''}
    Another participant suggested inferring user intents from the initial chart with errors or user modifications.
    
    \item {Consider semantic meaning.} 
    Some participants expected a more intelligent tool that can understand the semantics of the data field to better revise the chart. One participant mentioned,
    \textit{``I found that \name cannot detect whether the semantics in the chart is correct. It would be better if the tool could provide more revision suggestions based on the semantic meaning of data.''}
    
    \item {Add more visual principles.} Some participants suggested that more rules could be integrated into \name. For example, rules to detect missing values or outliers can be added to ensure the validity of the rendered visualization. Rules regarding visual aesthetics are also desired, \textit{``some visual guidelines on aesthetics could be added to help make the charts more visually pleasing.''} 
\end{itemize}


\section{Discussion}
In this section, we discuss the potential use scenarios of the \name framework, as well as current limitations to guide future research.

\subsection{Potential Use Scenarios}
\textbf{\name for other grammars.} In this work, we utilize Vega-Lite as the visualization specification syntax. In the future work, \name can incorporate other popular visualization grammars such as D3.js, ggplot2, and Matplotlib. One way of doing so is to translate onward-and-backward between Vega-Lite and other specification languages. Another solution is to implement various versions of \name respectively to suit different grammars.
In practice, a BI tool with over 100,000 users for the tech company Ant Group
has already adapted \name into its visualization library in Javascript\footnote{\url{https://ava.antv.vision/en/docs/guide/chart-linter/intro}} and configured its internal design guidelines into the linter
to meet user needs. This case proves the adaptability of \name in other programming languages.

\textbf{\name in BI tools.} The framework of \name can be applied in business intelligent tools. Although most BI tools provide drag-and-drop interactions for users to build visualizations without coding, some wrong actions may affect the correct representation of the information, such as mapping data to an inappropriate channel or applying the wrong data transformation. Embedding \name in BI tools could save time for digging into design guidelines, which is especially handy for inexperienced users who need to create visualizations quickly. The BI tool mentioned above has received positive feedback in practice, which enables users to find out optimal configurations quickly.
Furthermore, \name can serve as a collaborative visualization design system, ensuring that various visualizations built by different users in multiple views are consistent and compatible~\cite{8017651}. For example, linting tools could remind users to pay attention to set margins or encode with specific color palettes following the homogeneous design guidelines.

\textbf{\name for education.} During the interview, many participants felt surprised that \name helped them recognize design guidelines of visualization, some of which were overlooked occasionally. Consequently, one of the potential applications of \name is an auxiliary teaching tool for visualization education. Novices could use \name to verify whether the visualizations they create violate any design guidelines. During the trial-and-error process, one can obtain visualization knowledge gradually, which also echos the pedagogical value of ESLint~\cite{tomasdottir2018adoption}. 
\name can be valuable for studying visualization design guidelines, while instructors could also benefit from the automatically linting function to make quick judgment when checking assignments.

\subsection{Limitations and Future Work}
We summarize several limitations mentioned during interviews and found in the design and implementation process and propose future work directions. 

\underline{\textit{Explain the violated rules in more expressive ways.}} Currently, violated rules detected by the linter are only displayed in plain text. When interviewing the participants about the functionalities of \name, more details about the rules were reported as desired, such as in-depth descriptions and graphical demonstration of the rules. In the future, it is necessary to explain each rule in more detail, including complete documentation and examples of typical documents and don'ts illustrated with code and graphics. 

\underline{\textit{Extend the coverage of rule categories.}} All the four types of rules covered in \name focus on basic construction errors of visualizations. In the current stage, guidelines regarding perceptual expressiveness or visual encoding effectiveness have not been considered. Due to subjective considerations, there is no clear right or wrong judgment for such errors. Hence, more pilot studies should be conducted before applying subjective guidelines to \name. \rv{In the adapted version for the tech company mentioned earlier, some soft rules configured by the domain experts were applied. In the next step, we plan to collect and evaluate different categories of rules, covering expressiveness, aesthetic or stylistic issues, to enrich the capability of \name}.

\underline{\textit{Enrich system configurations and user interaction.}} The current design of the framework restrains the capability of error handling to the principles included in the rule base. However, in practice, it can be more helpful if the rule base allows extension. Therefore, in the next step, we consider making the rules user-configurable so that \name can fit in different scenarios. One implementation of \name is the web editor with linting and fixing functions, where limited interactions were provided. It is not intuitive enough for users to refer each rule to its corresponding coding snippets when the broken rules are listed in a separate panel. How to present the errors and the linkage among rules, codes, and visualization will be one of the future works of \name. 

\rv{\underline{\textit{Employ more state-of-art research to resolve performance issues.}} Currently, the scopes of design rules and actions in \name are relatively small, where input cases can be solved within seconds. It is undeniable that, with the progress of development, the ever-growing rule base and action space would increase the execution time of \name. Keep improving the framework for better performance utilizing state-or-art research~\cite{gebser2019multi, cuteri2020overcoming, morrison2016branch} is worth further studies.}

\section{Conclusion}
We present \name, a framework for automatically detecting flaws in visualizations and suggesting revisions, consisting of a visualization linter and a visualization fixer. The visualization linter employs design guidelines collected from prior research and incorporates Answer Set Programming to identify issues of the input visualizations. The visualization fixer resolves the detected violations by the linter and automatically suggests revisions by formulating the problem as a binary integer programming problem. 
A prototype of \name is implemented as an online Vega-Lite editor, showing how
developers can use \name to create visualization without flaws. An in-lab user study was also conducted to evaluate the effectiveness and efficiency of \name. The results showed that \name could help users identifying and fixing errors. In future work, we plan to extend the current framework with larger coverage of rules, more system configurations and user interactions, and to employ it in broader application scenarios.

\section*{Acknowledgments}
{This work was supported in part by the National Natural Science Foundation of China 62072338, 6200070909.}

\bibliographystyle{abbrv-doi}

\bibliography{main}
\end{document}